# ELECTROMAGNETIC OSCILLATIONS IN PERIODIC MEDIUMS OUTSIDE THE PASSBANDS

Ayzatsky M.I.[*], NSC KIPT, 1, Akademicheskay Str., Kharkov, 61108, Ukraine,
aizatsky@nik.kharkov.ua

**Abstract**
In this paper we represent the results of investigations of the electromagnetic waves in a layered dielectric that is limited from one side by a metal plate. There are electromagnetic oscillations, which evanesce in the direction of periodicity and propagate along the perpendicular direction, so they can be treated as surface waves. Existence of such waves gives possibility to create resonators filled by a layered dielectric with new type of eigen oscillations based on these evanescent waves.

## INTRODUCTION

It is well known [1], that in 1-D periodic mediums there are two different electromagnetic eigen-oscillations supported by medium without external currents and charges. In the certain frequency intervals (passbands) the electromagnetic oscillations represent wave process, which carry constant energy (in the case of absence of absorption) in direct or opposite directions. Between the passbands the electromagnetic oscillations have a structure that is distinct from previous case. In these frequency intervals electromagnetic oscillations transfer no energy in the direction of periodicity and have decreasing (increasing) dependence on the coordinate. These frequency intervals are called forbidden bands (stopbands). Today the 1-D periodic mediums are sometimes called 1-D photonic band gap (PBG) structures. Results of our investigations of the properties of the electromagnetic oscillations in the stopbands of a layered structure are represented. Particularly, we have shown that under some conditions the eigen-functions, which describe the field distributions, can have the zero values at definite planes, which are perpendicular to the direction of periodicity. Placing metal sheets in this points we can essentially change performances of the waves reflected from finite number of dielectric layers lying on a metal plate and create dielectric-metal resonance systems, which have oscillations with increasing or decreasing field distributions along the coordinate $z$. These phenomena can be treated in the terms of the so-called "defect modes" [2] as the "surface defect levels" that arise as the result of existence of the interface between the medium with a band structure and the medium that can be considered as an infinite potential wall. In this paper we represent the results of investigations of the evanescent in one direction and propagating in perpendicular one electromagnetic waves in the case when a layered structure is limited from one side by metal (or in the case when periodicity is reflected symmetrically relatively some plane). Existence of such waves gives possibility to create resonators filled by a layered dielectric with new type of eigen oscillations based on these evanescent waves.

## 1. MULTI-LAYER DIELECTRIC

Let us consider properties of eigen electromagnetic oscillations in a layered dielectric, which represents periodically repeating along the axes $z$ a set of layers with thickness $d_1$ and $d_2$ and with permittivities $\varepsilon_1$ end $\varepsilon_2$. In transversal directions ($x, y$) the layers are not limited. Dependence on time and the transversal coordinate $x$ we shall suppose as $\exp\{i\{k_x x - \omega t\}\}$.

Dependence of the transversal components of electromagnetic field on the longitudinal coordinate z can be found from the Maxwell equations and for arbitrary selected two adjoining layers ($i$=1,2) (from which we shall conduct numbering periods ($s$=0)) tangential electric field components can be written as:

$$E_{\tau,i}^{(0)} = E_i^{(+)} e^{ik_{i,z}z} + E_i^{(-)} e^{-ik_{i,z}z}, \qquad (1.1)$$

$$H_{\tau,i}^{(0)} = \frac{q_i}{Z_0}\left[ E_i^{(+)} e^{ik_{z,i}z} - E_i^{(-)} e^{-ik_{i,z}z} \right], \qquad (1.2)$$

where $k_{i,z} = \omega/c \times sqrt(\varepsilon_i - k_x^2 c^2/\omega^2)$, $E_i^{(\pm)}$ - constants and $q_i = \varepsilon_i / sqrt(\varepsilon_i - k_x^2 c^2/\omega^2)$ - for $p$-polarization and $q_i = -sqrt(\varepsilon_i - k_x^2 c^2/\omega^2)$ -for $s$-polarization

As the considered system is periodic along the axes z with period $D=d_1+d_2$, field components within the period with number $s$ can be determined by the expression:

$$E_{\tau,i}^{(s)} = \rho^s \left[ E_i^{(+)} e^{ik_{i,z}(z-sD)} + E_i^{(-)} e^{-ik_{i,z}(z-sD)} \right] \qquad (1.3)$$

where $\rho$ - some complex number. Similarly, one can write the expression for $H_{\tau,i}^{(s)}$. It is easy to show that at the fixed frequency the boundary conditions for the field components are fulfilled only for two values of parameter $\rho$ - $\rho_1$ and $\rho_2 = 1/\rho_1$:

$$\rho_{1,2} = Q \pm \sqrt{Q^2 - 1} \qquad (1.4)$$

where $Q$ is determined by the expression

$$Q = \frac{\alpha_1^2 \cos(k_{1,z}d_1 + k_{2,z}d_2) - \alpha_2^2 \cos(k_{1,z}d_1 - k_{2,z}d_2)}{\alpha_1^2 - \alpha_2^2}, \qquad (1.5)$$

and $\alpha_1 = (1+\sqrt{\varepsilon_2/\varepsilon_1})/2$, $\alpha_2 = (1-\sqrt{\varepsilon_2/\varepsilon_1})/2$.

Inside the passbans $|\rho_1| = |\rho_2| = 1$ and inside the stopbans $|\rho_1| < 1$ and $|\rho_2| > 1$ (see, Fig.1, where the dependence of $abs(\rho_1)$ on dimensionless frequency $\Omega = \omega D/(2\pi c) = D/\lambda$ is depicted for the case when $\kappa_x = k_x D/2\pi = D/\lambda_x = 0$ and $n_1 = sqrt(\varepsilon_1') = 2$,

---

[*] Aizatsky N.I.

$n_2 = sqrt(\varepsilon'_2) = 1.38$, $\varepsilon''_1 = \varepsilon''_2 = 0$, $d_1 = n_2/(n_1+n_2)D$, $d_2 = n_1/(n_1+n_2)D$. If $\kappa_x = k_x D/2\pi = D/\lambda_x > 0$, there is the gap for frequencies $0 < \Omega < \Omega_0$. This gap we shall call the natural or zero stop band.

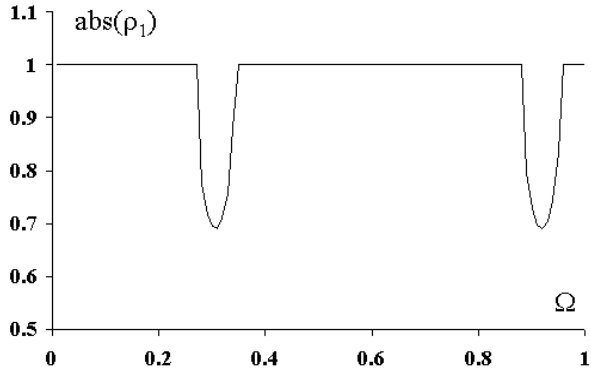

Figure 1

Let us consider a structure of an electrical field in some stop band. An example can be seen in Fig.2, where the dependencies of the modulus of $E_\tau$ on the longitudinal coordinate $\xi = z/D$ are shown within one period in the middle of the first stopband ($\Omega = 0.31$, see, Fig.1).

From Fig.2 it follows that both eigen non-propagating oscillations (d-decreasing - $|\rho_1| < 1$, i-increasing - $|\rho_2| > 1$) of layered dielectric have an interesting feature - in the certain planes, perpendicular to the $z$ axes, the tangential component of an electrical field equals to zero [3]. And for two fundamental solutions these planes do not coincide. Putting in these places metal planes, we can effectively control the field distribution.

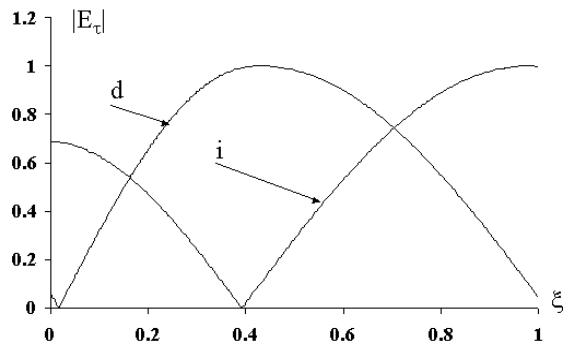

Figure 2

Indeed, a general solution of electrodynamic problem with a finite layered structure is represented as the sum of two fundamental (partial) solutions $E_\tau(z) = C_1 E_{\tau,1}(z) + C_2 E_{\tau,2}(z)$. If the metal plate is located at $z = z_*$, such condition must be fulfilled

$$C_1 E_{\tau,1}(z_*) + C_2 E_{\tau,2}(z_*) = 0. \quad (1.6)$$

From this condition it follows that if $E_{\tau,1}(z_*) = 0$ ($E_{\tau,2}(z_*) = 0$), then $C_2 = 0$ ($C_1 = 0$). Thus, choosing some $z_*$, we can create the field distribution that appropriate only to one fundamental solution, i.e. either pure increasing or decreasing one.

Transversal planes with zero value of the tangential component of electric field exist in all gaps under some conditions, but for the zero gap they locate near the edge of the gap and when $\kappa_x \to 0$ the frequency width of this interval tends to zero.

In section 2 we represent the results of exploring of the surface waves that can exist in the forbidden zones in the semi-limited layered dielectric, which ends by a metal plate.

Using the described circumstances, we can, at first, essentially change performances of waves reflected from finite number of dielectric layers, lying on a metal plate. Secondly, we can create metal-dielectric resonance systems with eigen-oscillations that is based on evanescent (non-propagating) waves and have increasing (decreasing) field distributions along the coordinate $z$. Results of these investigations is given in section 3.

## 2. SURFACE WAVES IN THE HALF-LIMITED LAYERED DIELECTRIC

Assume that the front surface of the metal plate has the coordinate $z=z_m$. The layered dielectric is spaced so that a layer with the higher refractive index starts at the coordinate $z=0$ (see Fig.3).

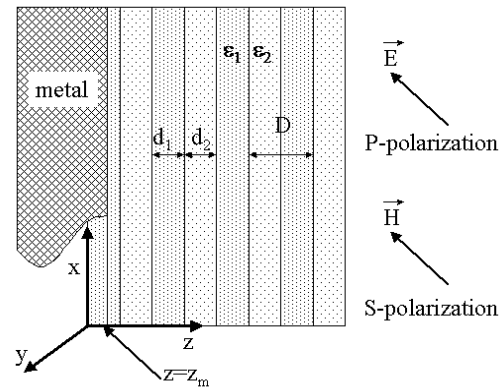

Figure 3

Changing the position of the front surface of the metal plate ($0<z_m<D$), we can study the different "surface levels" of the electromagnetic oscillations. We shall consider the case when optical length of each layer equals a quarter of some reference wavelength $\lambda_*$ - $d_i = \lambda_*/(4n_i)$. For ideal metal we have a boundary condition at $z=z_m$ that coincide with (1.6). If $E_{\tau,1}(z)$ describes the decreasing solution then we have to suppose that $C_2=0$ and try to find such frequency $\Omega$ and transversal wavenumber $\kappa_x$ inside a forbidden band when

$$E_{\tau,1}(\Omega, \kappa_x, z=z_m) = 0. \quad (1.7)$$

In Fig.4 (*p*-polarization) and Fig.5 (*s*-polarization) we represent the results of solving this dispersion equation (symbols {$\star$} - $z_m=0$, the layered medium starts with more dense layer - $n_1=2$; and symbols {+} - $z_m=d_1$, layered medium starts with a less dense layer - $n_2=1.38$). White color marks forbidden bands, grey color – passbands.

We can see that for *p*-polarization there are surface waves both for $z_m=0$ and $z_m=d_1$, but surface waves with *s*-polarization exist only when $z_m=0$. The surface wave in the zero forbidden band exists only for *p*-polarization when $0<z_m<d_1/2$ and $(d_1+d_2/2)<z_m<D$. In the first forbidden band for *p*-polarization waves can travel along the x-axis with phase velocities less or greater than that of light ($0<z_m<d_1/2$ and $(d_1+d_2/2)<z_m<D$) and they are slow waves when $d_1/2<z_m<(d_1+d_2/2)$.

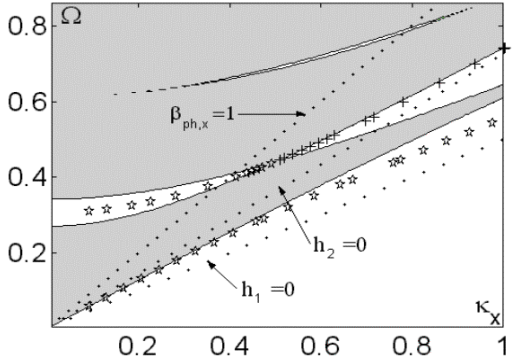

Figure 4

For *s*-polarization surface waves in the zero forbidden band cannot exist. In the first forbidden band waves travel along the x-axis with phase velocities less or greater than that of light only when ($0<z_m<d_1/2$ and $(d_1+d_2/2)<z_m<D$). So, surface waves with *s*-polarization do not exist when $d_1/2<z_m<(d_1+d_2/2)$.

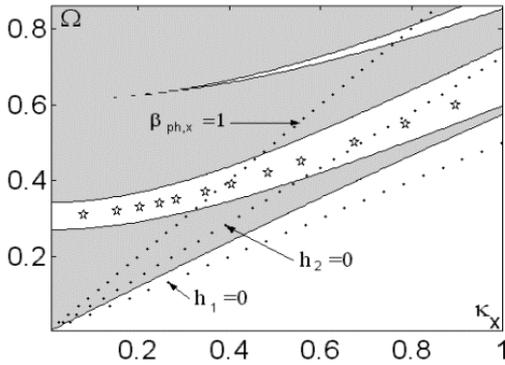

Figure 5

Our preliminary simulations have shown that studied surface waves can exist in the layered structures with finite number of layers limited from one side by metal.

## 3. NEW TYPE OF EIGEN OSCILLATIONS IN RESONATORS FILLED WITH A LAYERED DIELECTRIC

As the considered above system is periodic, then the condition (1.7) is fulfilled not only for $z=z_m$ but also for the set of points with longitudinal coordinates $z=z_m+D\times s$, $s=1,2,\ldots$. So, we can put an additional metal sheet in some point from this set and thus we get the waveguide with the non-uniform transverse distribution. All results, obtained in section 2, can be used for describing waves between such two metal sheets. We must note that in the case of two metal sheets there are two types of waves. The first type we shall obtain if we look at the system as beginning from one sheet (for example, $z_{m,1}=0$), and the second type - if we look at the system as beginning from another sheet (if $z_{m,1}=0$ then $z_{m,2}=d_1$).

We can fix the value of the transversal wavenumber $\kappa_x$ by restricting the layered dielectric along the transversal direction by the metal cylinder with radius *b* and in such way we obtain a closed cavity (see, Fig.6). For example, in the case of E-wave $\kappa_x=k_xD/2\pi=\lambda_{0p}D/(b2\pi)$, where $\lambda_{0p}$ are the roots of the Bessel function $J_0(x)$.

From Fig.4 it follows that for different values of $\kappa_x$ we may have either a decreasing (along the coordinate *z*) eigen oscillation ($\kappa_x<0.5$) or an increasing ($\kappa_x>0.5$) one.

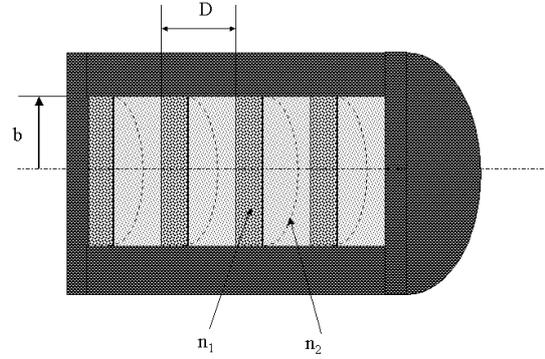

Figure 6

Using SUPERFISH code we simulated the properties of the resonator depicted on the Fig.6 ($n_1=2$, $n_2=1.38$, $D=2$ cm, $d_1=n_2/(n_1+n_2)D=0.816$ cm, $d_2=D-d_1$). For $b=4$ cm ($\kappa_x=0.191$) the value of eigen frequency equals $\Omega=0.3275$ and from Eq.(1.7) we obtained $\Omega=0.3274$. For this value of $\kappa_x$ the simulated longitudinal distribution of the electric field on the axis of resonator (*r*=0) is a decreasing one (see, Fig.7-1). For $b=1$ cm ($\kappa_x=0.7655$) the value of eigen frequency equals $\Omega=0.5905$ and from Eq.(1.7) we obtained $\Omega=0.591$. In this case the simulated longitudinal distribution of the electric field is an increasing one (see, Fig.7-2).

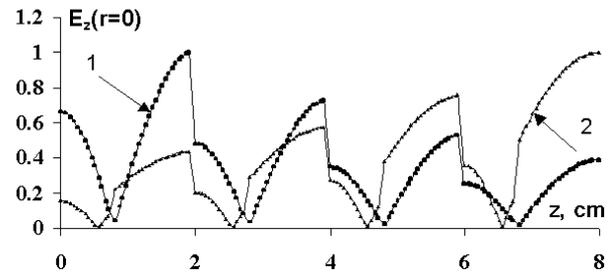

Figure 7

## SUMMARY

It was usually assumed that the resonator based on a waveguide has the eigen oscillations that are formed by interference of two waves which propagate in different directions and have equal amplitudes. These patterns are usually called standing waves. We have shown that the eigen oscillations of a resonator can be base on the evanescent (non-propagating) waves. In some cases we need only one eigen wave to compose the eigen oscillation of a closed cavity.